\documentclass[12pt,fleqn]{article}    
\usepackage{umlaute}
\usepackage{epsfig}
\begin{document}
\baselineskip 18pt  

\begin{center}
{\large {\bf Comment on "Forward K$^{+}$ - production in subthreshold \\
 pA collisions at 1.0 GeV" }}
\end{center}
\vskip 0.1cm
A Comment on the letter by V. Koptev et~al. - Phys. Rev. Lett. 87 (2001) 022301
\vskip 0.5cm

The recent paper by V. Koptev, M. B\"uscher et~al. (1) presents data
of great interest on the invariant cross section for the reaction
p(1.0 GeV/c) $^{12}$C $ \to $ K$^{+}$ X as a function of the K$^{+}$
momentum from 150 MeV/c to above 500 MeV/c (close to the exclusive
production of $^{12}_{\Lambda}$C). From the kinematics far below the
free NN threshold the data above 400 MeV/c should reflect a high
degree of collectivity among 6~-~7 bound nucleons following the
estimate of the authors. 

 \vskip 0.2cm

However, the conclusions above have to be taken with care: the
estimate lakes sufficient dynamical content assuming K$^{+}$
production in the collision of the projectile with a single heavy
nucleon  cluster. We approach the data differently;  experience from
other near-exclusive high momentum transfer processes suggest the
following microscopic picture: the dynamical collectivity (beyond the
pure center-of-mass correlation of all nucleons) is an interplay of
the strength of the NN rescattering amplitude, phase space and
momentum sharing constraints among the bound nucleons. At low K$^{+}$
momenta the characteristic momentum transfer of q $\sim \sqrt{3}$ M
$\sim $ 1500 MeV/c (M is the mass of the nucleon) is shared among
typically 4 nucleons: such a distribution combines maximal phase space
with momentum sharing on each bound state wave function near the peak
of the momentum distribution of a p-shell nucleon in $^{12}$C. With
the K$^{+}$ momentum above 450 MeV/c, phase space is dominated by
single nucleon emission, accompanied by the momentum sharing among
typically additional 2 bound nucleons (the nucleon momenta  exchanged
are close to the peak of the p-shell momentum density of around 200
MeV/c). A higher degree of collectivity is suppressed: little gain in
momentum sharing is overbalanced  by decreasing rescattering
contributions of higher order reflecting the strength of the
elementary (off-shell) NN amplitude. 

\vskip 0.1cm

We substantiate these heuristic arguments by a microscopic
calculation. We evaluate the full transition amplitude with N
collectively cooperating nucleons in $^{12}$C and M nucleons in the
continuum  
\begin{displaymath}
\hspace*{-1.0cm} T(N,M) = \sum_{i>j} < {\bf K}_{\Lambda}, {\bf K}_{1}\dots {\bf K}_{M} \, A(M + 1, \dots, N, \dots , 12 ) \left |
V (i,j, {\bf K}_{k} ) \right | \; {\bf K}, A(1, \dots , 12 )  >
\end{displaymath}
where we model $^{12}$C as s and p shell nucleons in an harmonic
oscillator potential (2). Along the same line we model the NN
rescattering amplitude as the exchange of effective ($ \pi \pi$) $
\sigma $ or $\rho $ mesons  including N$^{*}$ resonances in the mass
range of 1650 MeV up to 1800 MeV (3). For the evaluation we compare a
zero range approximation and a Gaussian parametrization of NN
interaction, where we fit the parameters to NN $ \to $ NN ($\pi \pi$)
data in the appropriate energy range (4); the meson NN couplings are
taken from the Bonn potential (5), while the couplings to the N*
resonances are extracted from the corresponding partial widths
(3). Implicitly we assume that K$^{+}$N rescattering is small: the
$\Lambda$K$^{+}$ final state is formed in the initial N$^{*}$
decay. The multi-nucleon phase space is included in a nonrelativistic
approximation (6). 

\vskip 0.1cm

We present the characteristic results of our calculation in Fig. 1: we
compare the invariant K$^{+}$ cross section for an increasing number
of active (collective) nucleons. The result qualitatively supports our
arguments above: summing up explicitly the influence of up to 6 bound
nuclears, we find for low K$^{+}$ momenta around 200 MeV/c dominantly
4 collectively interacting nucleons (whereby  typically 3 nucleons are
emitted into the continuum), while K$^{+}$ momenta around 500 MeV/c
involve 2~-~3 bound nucleons (with the knockout of a single nucleon).

\vskip 0.1cm

\begin{figure}[t]

\leftline{\hskip .8cm
\epsfig{file=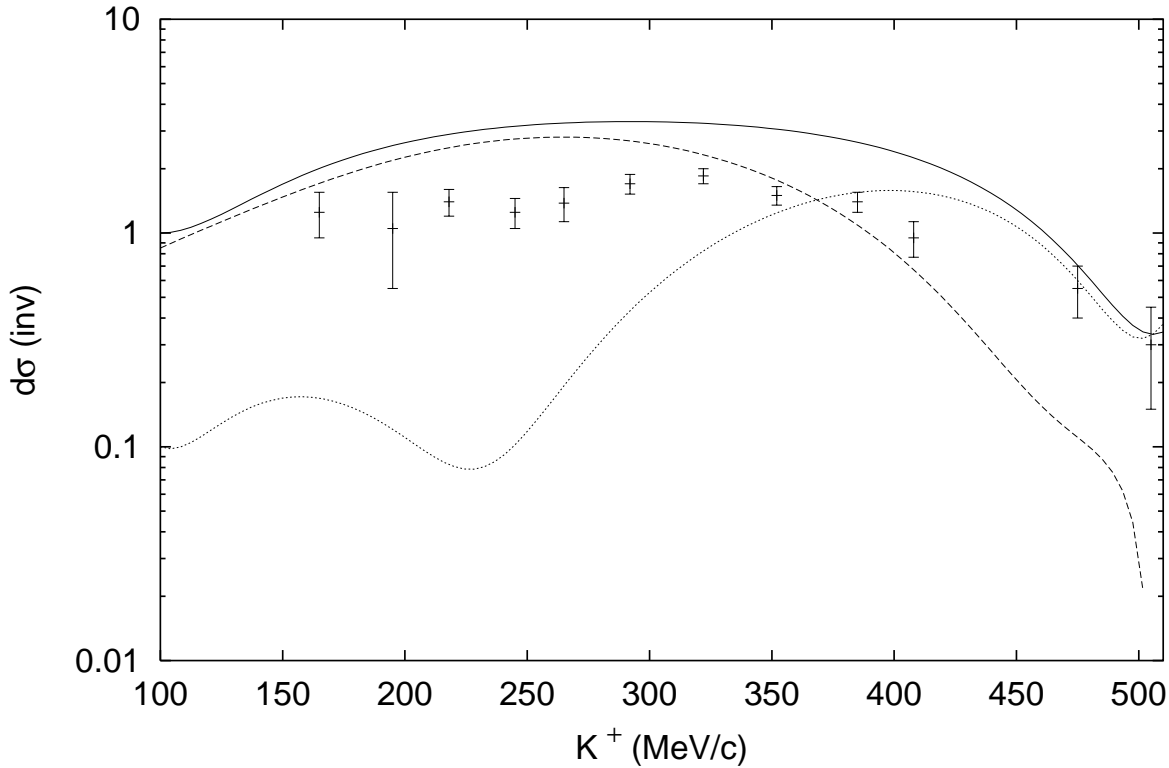,width=16cm}}

\caption{Invariant cross section for p$^{12}$C $\to $ K$^{+}$ X as a
function of the K$^{+}$ momentum. Compared are the full cross section
(solid line) with the contribution from $N=4$ and $N=2$ collective
nucleons (dashed and dotted line, respectively). The data shown are
from ref.~1.} 

\end{figure}

At present our picture is preliminary and has to be substantiated 
by further data (most interesting would be the transition to (nearly)
exclusive K$^{+}$ production in p-nucleus and in nucleus-nucleus
collisions, i.~e. complete K$^{+}$ fusion) and in further
calculations: so far our model in too crude for subtle information
like on the K$^{+}$ - self energy in nuclei (7). Beyond that, as the
energy sharing in collective phenomena is highly unknown, attempts
towards a covariant model calculation including full retardation,
which allows for a free energy and momentum sharing among the
interacting nucleons, would be highly desirable.  

\vskip 2.0cm

\baselineskip 14pt
M. Dillig \\
Institute for Theoretical Physics III \\
University of Erlangen-N\"urnberg \\
D-91058 Erlangen, Germany \\
email: mdillig@theorie3.physik.uni-erlangen.de
\vskip 0.2cm
PACS 13.60Le, 14.40 Cs 24.40-n \newline
\vskip 2.0cm
* Supported in part by the Kernforschungszentrum KFZ J\"ulich, Germany

\vskip 2.0cm

\begin{enumerate}
\baselineskip 14pt
\item V. Koptev, M. B\"uscher et al.: Phys. Rev. Lett. 87 (2001) 022301;
\item H. de Vries, C. W. de Jager and C. de Vries: At Data Nucl. Data
Tables 36 \\(1987) 495 
\item Rev. Part. Prop. (D.~E. Groom et~al.): Eur. Phys. Journ. C15 (2000) 1; 
\item Compilation of PP Cross Section: CERN-HERA 84-01 (1984); \\
   F. Smimizu et al.: Nucl. Phys. A 386 (1982) 571
\item Machleidt, R.: Adv. Nucl. Phys. 19 (1989) 189;
\item R. Shyam and J. Knoll: Nucl. Phys. A 426 (1984) 606;
\item G. Z. Rudy et~al.: nucl-th/0202069
\end{enumerate}

\end{document}